\documentclass[12pt]{article}
\DeclareUnicodeCharacter{2212}{-}
\usepackage{url}
\usepackage{listings,jvlisting}
\usepackage{amssymb}
\usepackage{tabularx}
\usepackage{amsmath}
\usepackage{mdframed}
\usepackage[pdftex]{graphicx}
\usepackage{setspace}
\usepackage{here}
\usepackage{authblk}
\usepackage{mathtools}
\usepackage{amsthm}
\usepackage{algorithmic}
\usepackage{algorithm}
\makeatletter

\@addtoreset{equation}{section}
\makeatother

\makeatletter
\newcommand{\figcaption}[1]{\def\@captype{figure}\caption{#1}}
\newcommand{\tblcaption}[1]{\def\@captype{table}\caption{#1}}
\makeatother

\def\ep{{\varepsilon}}
\def\la{{\lambda}}

\def\kh{{\hat k}}

\def\ph{{\hat p}}

\def\[{{\text{\boldmath $[$}}}
\def\]{{\text{\boldmath $]$}}}

\def\|{{\,|\,}}

\def\/{{\Bigr/\!\!}}

\def\1r{{\rm (1)}}
\def\2r{{\rm (2)}}
\def\3r{{\rm (3)}}
\def\4r{{\rm (4)}}
\def\5r{{\rm (5)}}

\begin{document}
\title{Application of Multi-Armed Bandits to Model-assisted designs for Dose-Finding Clinical Trials}
\author[1,2]{Masahiro Kojima\footnote{Address: Biometrics Department, R\&D Division, Kyowa Kirin Co., Ltd.
Otemachi Financial City Grand Cube, 1-9-2 Otemachi, Chiyoda-ku, Tokyo, 100-004, Japan. Tel: +81-3-5205-7200 \quad
E-Mail: masahiro.kojima.tk@kyowakirin.com}}
\affil[1]{Kyowa Kirin Co., Ltd}
\affil[2]{The Graduate University for Advanced Studies}

\maketitle

\abstract{\noindent
We consider applying multi-armed bandits to model-assisted designs for dose-finding clinical trials. Multi-armed bandits are very simple and powerful methods to determine actions to maximize a reward in a limited number of trials. Among the multi-armed bandits, we first consider the use of Thompson sampling which determines actions based on random samples from a posterior distribution. In the small sample size, as shown in dose-finding trials, because the tails of posterior distribution are heavier and random samples are too much variability, we also consider an application of regularized Thompson sampling and greedy algorithm. The greedy algorithm determines a dose based on a posterior mean. In addition, we also propose a method to determine a dose based on a posterior median. We evaluate the performance of our proposed designs for six scenarios via simulation studies.
}
\par\vspace{4mm}

\section{Introduction}
\label{sec1}

The primary objective of model-assisted designs for dose-finding trials is to identify a maximum tolerated dose (MTD) in terms of safety. The dose-finding trials are conducted on a limited small sample size, starting with a lowest dose and repeating a dose-assignment to identify a dose near a target toxicity level. Designs for a dose-assignment have been often used the 3+3 design \cite{Storer1989} and continual reassessment method (CRM) \cite{OQuigley1990}. Recently, model-assisted designs which are no complex assumptions and have superior performance have been proposed. The model-assisted designs include Bayesian optimal interval (BOIN) design \cite{Liu2015}, keyboard design \cite{Yan2017}, and related designs\cite{Lee2019,Lin2020,Kojima2021a,Kojima2021b,Kojima2021c,Kojima2021d,Mozgunov2020a,Mozgunov2020b,Lin2020b}. The BOIN design conducts a dose-assignment by using an interval for the toxicity rate that is optimized by minimizing errors in dose-assignment decisions. The Keyboard design divides the beta posterior distribution of the toxicity rate and adjusts the dose based on the interval with the maximum interval probability.

Multi-armed bandits are very simple and powerful methods to determine actions to maximize a reward in a limited number of trials. The origin of the term "multi-armed bandit" comes from the scenario that a gambler challenges multiple slot machines (one-armed bandits) with different payoffs in order to maximize the reward. The features of multi-armed bandits are not only to explore effective actions, but also to exploit more effective actions. Multi-armed bandits were originally proposed for a phase III clinical trial with one control treatment vs one alternative \cite{Thompson1933,Gittins1979}. A phase II/III seamless trial design is a kind of multi-armed bandits design because the effective dose is identified in a phase II, and then the number of patients treated identified dose is increased in phase III to verify the efficacy. Aziz et al. \cite{Aziz2021} proposed multi-armed bandit designs based on a CRM, which showed a higher percentage of correct MTD identification than a CRM design. However, because the CRM requires complex assumptions of a statistical model at advance, the multi-armed bandit designs also require complex assumptions.

In this paper, we propose novel multi-armed bandit BOIN and keyboard designs. Among methods of multi-armed bandits, we first consider the use of Thompson sampling which determines a dose based on random samples from a posterior distribution. In the small sample size, as shown in dose-finding trials, because the tails of posterior distribution are heavier and random samples are too much variability, we also consider an application of regularized Thompson sampling and greedy algorithm. The greedy algorithm determines a dose based on a posterior mean. In addition, we also propose a method to determine a dose based on the posterior median. In simulation studies, we evaluate a performance of the multi-armed bandit BOIN and keyboard designs for six scenarios.

The paper is structured as follows: Section 2 presents multi-armed bandit BOIN and keyboard designs. In addition, we present a simulation configuration to confirm a performance for the multi-armed bandit BOIN and keyboard designs. Section 3 presents the simulation results. Section 4 discusses our proposed designs and the results.

\section{Methods}\label{sec2}
We consider a dose-finding trial with $K$ doses and sample size $N$ for identifying a maximum tolerated dose (MTD). We evaluate the safety of drugs per a cohort. The cohort size is usually three. We assume that a target toxicity level (TTL) is $\phi$, the number of patients treated at the current dose $k$ is $n_k$, and the number of DLTs at the current dose is $m_k$. The MTD is the dose whose the observed toxicity rate ($\ph_k=\frac{m_k}{n_k}$) is closest to $\phi$ after all $N$ patients have been treated. From a safety perspective, if $Pr(p_k>\phi|n_k,m_k)>0.95$ holds, the dose $k$ and higher doses are excluded in the dose-finding trial. The exclusion of the dose is called a stopping rule. First of all, we review BOIN and keyboard designs.

\subsection{Review of Bayesian Optimal Interval and Keyboard Designs}
{\bf Bayesian optimal interval (BOIN) design.} The BOIN design conducts a dose-assignment according to a dose escalation boundary $\la_e(\phi)$ and de-escalation boundary $\la_d(\phi)$. The dose-assignment rules are, if $\ph_k\in\left(\la_e(\phi),\la_d(\phi)\right)$, to retain the current dose level for the next dose, if $\ph_k$ is above $\la_d(\phi)$, to de-escalate the next dose level, if $\ph_k$ is below $\la_e(\phi)$, to escalate the next dose level. For example, when $\phi$ is 0.3, the dose retainment interval is $(0.236, 0.358)$. 

\noindent
{\bf Keyboard design.} The keyboard design is a little more complex than the BOIN design, but has high identification rate of correct MTD as well as the BOIN design. In the keyboard design, $m_k$ is distributed according to the binomial distribution Binom$(n_k,p_k)$, the prior distribution of $p_k$ is the beta distribution Beta$(1,1)$, and the posterior distribution of $p_k$ is the Beta$(m_k+1,n_k-m_k+1)$. Because the target dose is the dose at which the observed toxicity rate is close to the TTL ($\phi$), the interval $(\phi-0.05,\phi+0.05)$ is used as the target key. The left and right sides of the target key are divided by a width of 0.1. For the posterior beta distribution of $p_k$, when the interval probability of the target key is the highest compared to other interval probabilities, the next dose level is retained the current dose level. If the interval probability of a key with a lower toxicity rate is the highest, the next dose level escalates. If the interval probability of a key with a higher toxicity rate is the highest, the next dose level de-escalates. 

In the following section, we propose applying multi-armed bandits to BOIN and Keyboard designs.

\subsection{Multi-Armed Bandit Bayesian Optimal Interval Design}
We consider combining the dose retainment interval of BOIN design with Thompson sampling. Thompson sampling selects an action that yields the largest reward with respect to random samples from a posterior distribution. For the dose-finding trials, because the largest reward is the toxicity rate close to the TTL, we propose the following Algorithm \ref{alg1}. $k_{max}$ is a maximum dose level which has been administered.

\begin{figure}[h!]
\begin{center}
\begin{minipage}{14cm}
\begin{algorithm}[H]
    \caption{BOIN design using Thompson sampling}
    \label{alg1}
    \begin{algorithmic}[1]
    \FOR{cohort$=1,2,\dots$}
    \FOR{$k=1,\dots, k_{max}$}
    \STATE sample $p^{\ast}_k\sim\mbox{Beta}(m_k+1,n_k-m_k+1)$
    \ENDFOR
    \IF{There is $p^{\ast}_k\in\left(\la_e(\phi),\la_d(\phi)\right)$}
    \STATE $\kh\leftarrow \underset{k:p^{\ast}_k\in\left(\la_e(\phi),\la_d(\phi)\right)}{\mbox{argmax}}(p^{\ast}_k)$
    \ELSIF{There is $p^{\ast}_k\leq\la_e(\phi)$}
    \STATE $\kh\leftarrow \underset{k:p^{\ast}_k\leq\la_e(\phi)}{\mbox{argmax}}(p^{\ast}_k)$
    \ELSIF{There is $p^{\ast}_k\geq\la_d(\phi)$}
    \STATE $\kh\leftarrow \underset{k:p^{\ast}_k\geq\la_d(\phi)}{\mbox{argmin}}(p^{\ast}_k)$
    \ENDIF
    \STATE treat $\kh$ dose level for the next dose
    \ENDFOR
    \end{algorithmic}
\end{algorithm}
\end{minipage}
\end{center}
\end{figure}

The line 3 extracts randomly one sample $p^\ast_k$ from the posterior beta distribution for each dose, respectively. We are interested in doses with a toxicity rate near the TTL. In the line 5, we first check whether or not there is $p^\ast_k$ for all $k$ within the dose-retainment interval around the TTL, and if there is, we hold the maximum dose level of $p^\ast_k$ within the dose-retainment interval at $\kh$. If there is no $p^\ast_k$ for all $k$ within the dose-retainment interval, then for doses with a lower toxicity rate than the boundary for the dose escalation, the next dose level escalates to one dose level above the administered maximum dose level. If there is only doses with a higher toxicity rate than the boundary for the dose de-escalation, the next dose level de-escalates to one dose level below the administered maximum dose level.

The dose-finding trials are often conducted with a small sample size of about 20 to 40. Because the tails of posterior beta distribution are heavier when the observed data is a little, the random samples from the posterior beta distribution on Thompson sampling are too much variability. To reduce the variability, we consider regularized Thompson sampling-$\ep$ and greedy algorithm. Thompson sampling-$\ep$ re-samples $p^{\ast}_k$ if $p^{\ast}_k$ is not in $\[\ph_k-\ep,\ph_k+\ep\]$ for each $k$. The algorithm of Thompson sampling-$\ep$ only needs to modify the line 3 of Algorithm \ref{alg1}. In the subsequent methods, we only need to modify the line 3. The greedy algorithm deals with the posterior mean instead of the random sample from posterior distribution, the code of line 3 is $p^{\ast}_k\leftarrow\frac{m_k+1}{n_k+2}$. In addition, we consider the observed toxicity rate $\ph_k$ from frequentist perspective. When the prior distribution of $p_k$ is Beta$(1,1)$, the posterior median becomes $\ph_k$. The algorithm that replaces the lines as $p^{\ast}_k\leftarrow\frac{m_k}{n_k}$ is called the median algorithm in this paper.

\subsection{Multi-Armed Bandit Keyboard Design}
First of all, we define the notations of the keys of keyboard design. Let the target key be $\mbox{key}_t(\phi)=(\phi-0.05,\phi+0.05)$, the left side of the target key be $\mbox{key}_1(\phi),\ldots,\mbox{key}_{t-1}(\phi)$, the right side be $\mbox{key}_{t+1}(\phi),\ldots$.

\begin{figure}[h!]
\begin{center}
\begin{minipage}{14cm}
\begin{algorithm}[H]
    \caption{Keyboard design using Thompson sampling}
    \label{alg2}
    \begin{algorithmic}[1]
    \FOR{cohort$=1,2,\dots$}
    \FOR{$k=1,\dots, k_{max}$}
    \STATE sample $p^{\ast}_k\sim\mbox{Beta}(m_k+1,n_k-m_k+1)$
    \ENDFOR
    \IF{There is $p^{\ast}_k\in \text{key}_t(\phi)$}
    \STATE $\kh\leftarrow \underset{k:p^{\ast}_k\in \text{key}_t(\phi)}{\mbox{argmax}}(p^{\ast}_k)$
    \ELSIF{There is $p^{\ast}_k \in \{\mbox{key}_1(\phi),\ldots,\mbox{key}_{t-1}(\phi)\}$}
    \STATE $\kh\leftarrow \underset{k:p^{\ast}_k\in \left\{\text{key}_1(\phi),\ldots,\text{key}_{t-1}(\phi)\right\}}{\mbox{argmax}}(p^{\ast}_k)$
    \ELSIF{There is $p^{\ast}_k \in \{\mbox{key}_{t+1}(\phi),\ldots\}$}
    \STATE $\kh\leftarrow \underset{k:p^{\ast}_k\in\left\{\text{key}_{t+1}(\phi),\ldots\right\}}{\mbox{argmin}}(p^{\ast}_k)$
    \ENDIF
    \STATE treat $\kh$ dose level for the next dose
    \ENDFOR
    \end{algorithmic}
\end{algorithm}
\end{minipage}
\end{center}
\end{figure}

In the lines 5-6, the dose level of the highest $p^\ast_k$ within the target key is prescribed for the next cohort. If there is no $p^\ast_k$ within the target key, then the next dose is one higher dose level that is the highest $p^\ast_k$ within $\{\mbox{key}_1(\phi),\ldots,\mbox{key}_{t-1}(\phi)\}$. In the lines 9-10, the next dose is one higher dose level that is the smallest $p^\ast_k$ within $\{\mbox{key}_{t+1},\ldots\}$.

\subsection{Simulation Configuration}
We perform simulation studies to confirm the performance of multi-armed bandit BOIN and keyboard designs compared to the BOIN and keyboard designs. For the multi-armed bandit designs, we prepare Thompson sampling (TS), Thompson sampling-$\ep(=$$0.05)$ (TS5\%), greedy algorithm (G), and median algorithm (M). We assume that the sample size is 36, the cohort size is 3, the dose level is 6, and the TTL is 30\%. The number of simulation times was 10,000. We prepare six scenarios in which the correct MTDs are arranged in order from the maximum dose level 6 to the minimum dose level 1. The detailed toxicity rates for each scenario are shown in Table \ref{Table1} and \ref{Table2}. We evaluate the percentage of correct MTD identification and the average number of patients treated on the correct MTD.

\section{Results}\label{sec3}
\begin{figure}[h!]
  \begin{center}
  \includegraphics[width=15cm]{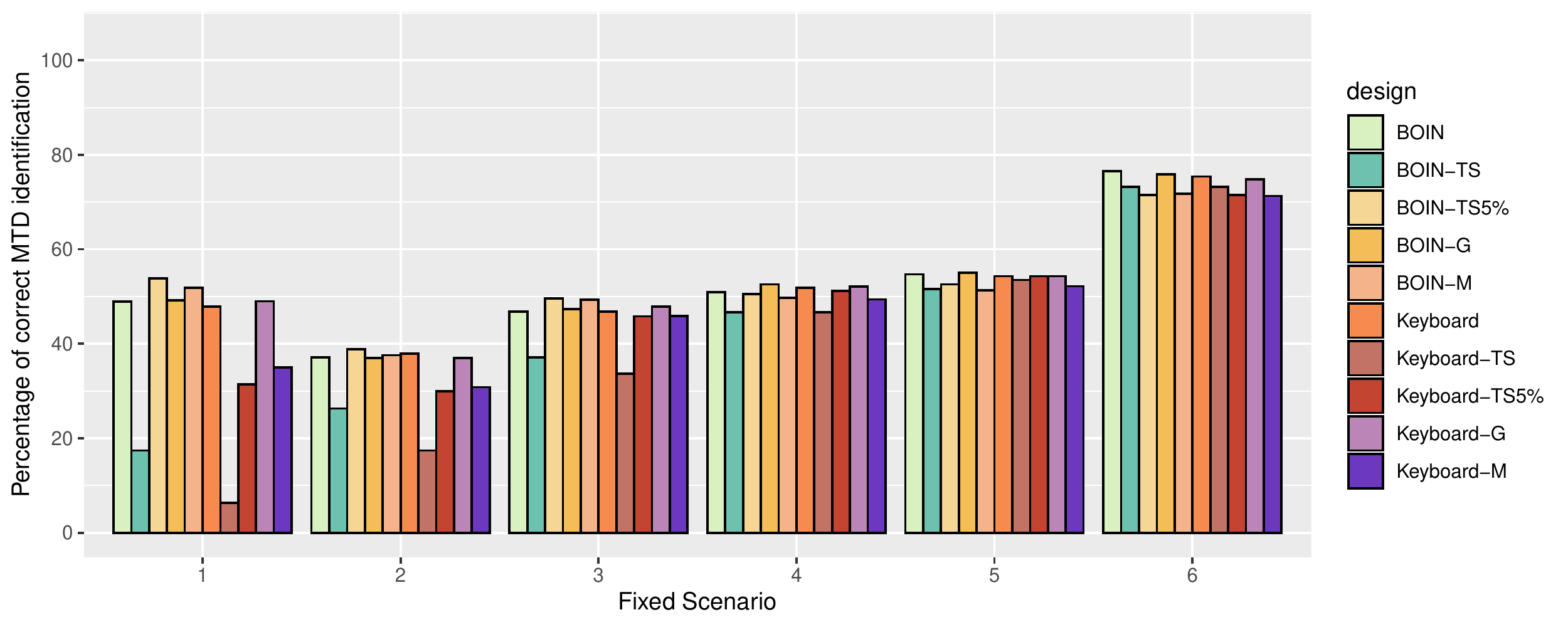}
  \\
      \footnotesize{TS: Thompson sampling. TS5\%: Thompson sampling-0.05. G: Greedy algorithm. M: Median algorithm}\\
        \caption{Simulation results for correct MTD identification\label{Figure1}}
  \end{center}
\end{figure}

We show the results of the simulation studies. Figure \ref{Figure1} shows the percentages of correct MTD identification (PCMI) for all scenarios. The PCMIs for BOIN-TS and Keyboard-TS did not exceed the PCMIs for BOIN and Keyboard designs in any scenario. The PCMIs for BOIN-TS5\% outperformed the PCMIs for BOIN design in scenarios 1, 2, and 3. However, the PCMI for BOIN-TS5\% was lower than BOIN design when the true MTD was dose levels 1-3. The PCMIs for BOIN-G had higher than the PCMIs for BOIN in scenarios 1, 3, 4, and 5. The PCMIs for BOIN-M had higher than the PCMIs for BOIN in scenarios 1, 2, and 3. For the multi-armed bandit keyboard designs, the PCMIs for keyboard-G had higher than the PCMIs for keyboard in scenarios 1, 3, and 4. However, the other multi-armed bandit keyboard designs did not have higher PCMIs for keyboard design in any scenario. The detailed numerical results and the number of patients treated for each dose are shown in Tables \ref{Table1} and \ref{Table2}. We show the results of PCMI for different values of $\ep$ in Thompson sampling-$\ep$ in Supplemental Table \ref{STable1}. The results were stable when $\ep$ was less than 5\%.

\begin{table}[h!]
  \begin{center}
\caption{Simulation results for MTD identification (scenarios 1-3)\label{Table1}}
{\tabcolsep=4.25pt
\begin{tabular}{|c|c|c|c|c|c|c||c|c|c|c|c|c|}\hline
   & \multicolumn{6}{c||}{\%MTD identification}&\multicolumn{6}{c|}{\#patients treated}\\\hline
Design& 1 & 2 & 3 & 4 & 5 & 6& 1 & 2 & 3 & 4 & 5 & 6 \\\hline
{\it Scenario 1} ({\it \%Tox}) & {\it 5} & {\it 6} & {\it 8} & {\it 11} & {\it 19} & \underline{\it 32} & {\it 5} & {\it 6} & {\it 8} & {\it 11} & {\it 19} & \underline{\it 32} \\\hline
BOIN&2.1&3.1&5.1&10.2&30.6&\underline{48.9}&3.6&3.8&4.2&5.5&8.5&\underline{10.3}\\\hline
BOIN-TS&5.8&9.1&12.4&23.3&32.0&\underline{17.4}&7.5&7.4&7.3&6.7&4.9&\underline{2.3}\\\hline
BOIN-TS5\% &1.3&2.3&4.0&9.4&29.1&\underline{\bf 53.8}&3.6&3.8&4.2&5.2&7.8&\underline{\bf 11.4}\\\hline
BOIN-G&1.6&2.5&4.7&10.4&31.6&\underline{\bf 49.2}&4.1&4.3&4.9&6.0&8.2&\underline{8.5}\\\hline
BOIN-M&2.0&3.2&4.6&10.1&28.2&\underline{\bf 51.9}&3.6&3.7&4.1&5.0&7.7&\underline{\bf 11.9}\\\hline
Keyboard&2.0&3.1&5.3&10.2&31.6&\underline{47.9}&3.6&3.8&4.3&5.5&8.5&\underline{10.3}\\\hline
Keyboard-TS&9.0&13.3&19.6&27.8&24.0&\underline{6.3}&7.3&10.0&8.6&6.2&3.1&\underline{0.8}\\\hline
Keyboard-TS5\% &10.1&10.2&11.5&12.0&24.9&\underline{31.4}&3.6&6.3&6.4&6.2&6.6&\underline{6.8}\\\hline
Keyboard-G&1.5&2.5&4.3&10.0&32.7&\underline{\bf 49.0}&4.1&4.3&4.8&5.9&8.4&\underline{8.5}\\\hline
Keyboard-M&9.7&10.9&10.8&10.7&22.9&\underline{35.0}&3.6&6.2&6.2&6.0&6.5&\underline{7.6}\\\hline\hline
{\it Scenario 2} ({\it \%Tox}) & {\it 6} & {\it 8} & {\it 12} & {\it 18} & \underline{\it 30} & {\it 41}   & {\it 6} & {\it 8} & {\it 12} & {\it 18} & \underline{\it 30} & {\it 41}\\\hline
BOIN&2.6&4.5&10.5&27.1&\underline{37.1}&18.2&3.8&4.4&5.8&8.9&\underline{8.6}&4.6\\\hline
BOIN-TS&5.8&10.0&19.3&33.0&\underline{26.3}&5.6&7.7&7.9&8.1&7.1&\underline{4.0}&1.2\\\hline
BOIN-TS5\%&1.9&4.2&8.9&24.9&\underline{\bf 38.8}&21.3&3.9&4.4&5.5&7.9&\underline{\bf 8.9}&5.4\\\hline
BOIN-G&1.9&4.4&11.2&28.3&\underline{37.0}&17.2&4.3&5.0&6.5&8.7&\underline{8.1}&3.3\\\hline
BOIN-M&2.6&4.5&9.4&23.8&\underline{\bf 37.6}&22.1&3.7&4.1&5.2&7.8&\underline{\bf 9.2}&6.0\\\hline
Keyboard&2.9&4.9&10.0&26.5&\underline{37.9}&17.8&3.8&4.3&5.8&8.8&\underline{8.6}&4.7\\\hline
Keyboard-TS&7.8&12.8&24.4&35.0&\underline{17.4}&2.5&7.5&10.4&9.1&6.2&\underline{2.4}&0.5\\\hline
Keyboard-TS5\%&9.0&9.9&14.3&25.7&\underline{30.0}&11.1&3.8&6.7&7.6&8.1&\underline{6.8}&3.0\\\hline
Keyboard-G&2.1&4.1&10.4&28.6&\underline{37.0}&17.8&4.3&5.0&6.5&8.7&\underline{8.0}&3.4\\\hline
Keyboard-M&10.1&11.0&12.3&22.4&\underline{30.8}&13.4&3.7&6.7&7.0&7.9&\underline{7.2}&3.4\\\hline\hline
{\it Scenario 3} ({\it \%Tox}) & {\it 5} & {\it 10} & {\it 20} & \underline{\it 29} & {\it 50} & {\it 70} & {\it 5} & {\it 10} & {\it 20} & \underline{\it 29} & {\it 50} & {\it 70} \\\hline
BOIN&2.5&9.3&31.6&\underline{46.8}&9.7&0.1&3.8&6.0&10.3&\underline{11.2}&4.3&0.4\\\hline
BOIN-TS&3.6&12.7&38.0&\underline{37.1}&8.3&0.2&7.6&8.8&9.8&\underline{7.0}&2.5&0.3\\\hline
BOIN-TS5\%&1.9&8.3&27.8&\underline{\bf 49.6}&12.1&0.3&3.7&5.5&9.5&\underline{\bf 11.7}&5.2&0.5\\\hline
BOIN-G&2.0&8.1&33.0&\underline{\bf 47.3}&9.2&0.3&4.1&6.2&10.9&\underline{10.9}&3.6&0.3\\\hline
BOIN-M&2.4&8.6&27.6&\underline{\bf 49.3}&11.9&0.2&3.6&5.1&9.2&\underline{\bf 11.8}&5.6&0.6\\\hline
Keyboard&2.7&9.0&31.1&\underline{46.8}&10.2&0.3&3.7&6.0&10.3&\underline{11.2}&4.3&0.4\\\hline
Keyboard-TS&5.0&13.9&41.3&\underline{33.7}&6.0&0.1&7.3&11.1&10.1&\underline{5.8}&1.6&0.1\\\hline
Keyboard-TS5\%&5.9&9.5&30.1&\underline{45.8}&8.6&0.1&3.6&7.1&10.4&\underline{10.8}&3.8&0.3\\\hline
Keyboard-G&1.8&8.7&32.1&\underline{\bf 47.9}&9.2&0.2&4.1&6.2&10.9&\underline{11.0}&3.6&0.3\\\hline
Keyboard-M&6.9&9.9&27.3&\underline{45.9}&10.0&0.1&3.6&6.9&9.9&\underline{11.1}&4.2&0.3\\\hline
\end{tabular}}
  \end{center}
\end{table}

\begin{table}[h!]
  \begin{center}
\caption{Simulation results for MTD identification (scenarios 4-6)\label{Table2}}
{\tabcolsep=4.25pt
\begin{tabular}{|c|c|c|c|c|c|c||c|c|c|c|c|c|}\hline
   & \multicolumn{6}{c||}{\%MTD identification}&\multicolumn{6}{c|}{\#patients treated}\\\hline
Design& 1 & 2 & 3 & 4 & 5 & 6& 1 & 2 & 3 & 4 & 5 & 6 \\\hline
{\it Scenario 4} ({\it \%Tox}) & {\it 8} & {\it 15} & \underline{\it 29} & {\it 43} & {\it 50} & {\it 57} & {\it 8} & {\it 15} & \underline{\it 29} & {\it 43} & {\it 50} & {\it 57}  \\\hline
BOIN&6.1&23.3&\underline{50.9}&16.4&3.0&0.3&4.7&10.1&\underline{13.5}&6.1&1.3&0.2\\\hline
BOIN-TS&7.5&27.0&\underline{46.7}&15.9&2.7&0.2&8.7&10.9&\underline{10.6}&4.6&1.0&0.1\\\hline
BOIN-TS5\%&5.1&21.0&\underline{50.5}&19.5&3.6&0.3&4.5&8.6&\underline{\bf 13.8}&7.2&1.6&0.2\\\hline
BOIN-G&4.9&23.3&\underline{\bf 52.6}&16.5&2.4&0.2&5.2&9.9&\underline{\bf 14.2}&5.8&0.9&0.1\\\hline
BOIN-M&5.5&21.2&\underline{49.7}&19.5&3.7&0.4&4.3&8.2&\underline{\bf 13.8}&7.6&1.8&0.3\\\hline
Keyboard&5.9&22.5&\underline{51.9}&16.6&2.7&0.3&4.7&10.0&\underline{13.6}&6.1&1.3&0.2\\\hline
Keyboard-TS&8.3&28.3&\underline{46.7}&14.8&1.8&0.1&8.4&12.9&\underline{10.1}&3.8&0.7&0.1\\\hline
Keyboard-TS5\%&7.2&22.1&\underline{51.2}&17.0&2.3&0.1&4.4&10.3&\underline{\bf 13.9}&6.2&1.0&0.1\\\hline
Keyboard-G&4.5&24.1&\underline{\bf 52.1}&16.9&2.2&0.2&5.1&10.0&\underline{\bf 14.1}&5.8&0.9&0.1\\\hline
Keyboard-M&7.9&21.9&\underline{49.4}&17.3&3.2&0.3&4.2&10.0&\underline{\bf 13.8}&6.6&1.3&0.1\\\hline\hline
{\it Scenario 5} ({\it \%Tox}) & {\it 13} & \underline{\it 28} & {\it 41} & {\it 50} & \underline{\it 60} & {\it 70} & {\it 13} & \underline{\it 28} & {\it 41} & {\it 50} & \underline{\it 60} & {\it 70} \\\hline
BOIN&20.9&\underline{54.7}&20.8&3.4&0.2&0.0&10.2&\underline{16.0}&7.7&1.8&0.3&0.0\\\hline
BOIN-TS&22.9&\underline{51.6}&21.2&3.9&0.3&0.0&12.4&\underline{14.2}&7.1&2.0&0.3&0.0\\\hline
BOIN-TS5\%&19.2&\underline{52.6}&23.8&4.1&0.3&0.0&8.3&\underline{16.0}&9.2&2.2&0.3&0.0\\\hline
BOIN-G&20.1&\underline{\bf 55.0}&22.4&2.5&0.2&0.0&9.5&\underline{\bf 17.1}&7.8&1.4&0.1&0.0\\\hline
BOIN-M&19.9&\underline{51.3}&24.4&4.1&0.3&0.0&7.9&\underline{15.8}&9.5&2.4&0.3&0.0\\\hline
Keyboard&20.7&\underline{54.3}&21.6&3.1&0.2&0.0&10.2&\underline{15.8}&7.9&1.8&0.2&0.0\\\hline
Keyboard-TS&22.4&\underline{53.5}&20.4&3.4&0.3&0.0&12.0&\underline{15.2}&6.8&1.7&0.3&0.0\\\hline
Keyboard-TS5\%&18.4&\underline{54.3}&23.8&3.3&0.2&0.0&8.2&\underline{\bf16.7}&9.2&1.8&0.2&0.0\\\hline
Keyboard-G&20.5&\underline{54.3}&22.3&2.8&0.2&0.0&9.6&\underline{\bf 17.1}&7.8&1.4&0.1&0.0\\\hline
Keyboard-M&19.3&\underline{52.2}&24.5&3.8&0.2&0.0&8.0&\underline{\bf 16.4}&9.3&2.1&0.2&0.0\\\hline\hline
{\it Scenario 6} ({\it \%Tox}) & \underline{\it 28} & {\it 42} & {\it 49} & {\it 61} & {\it 76} & {\it 87} & \underline{\it 28} & {\it 42} & {\it 49} & {\it 61} & {\it 76} & {\it 87}  \\\hline
BOIN&\underline{76.5}&20.5&2.8&0.2&0.0&0.0&\underline{24.8}&9.0&2.0&0.3&0.0&0.0\\\hline
BOIN-TS&\underline{73.2}&21.6&4.6&0.5&0.0&0.0&\underline{22.7}&9.7&3.0&0.6&0.0&0.0\\\hline
BOIN-TS5\%&\underline{71.5}&24.0&4.3&0.2&0.0&0.0&\underline{22.1}&10.8&2.7&0.4&0.0&0.0\\\hline
BOIN-G&\underline{75.9}&21.0&3.0&0.1&0.0&0.0&\underline{24.7}&9.4&1.7&0.2&0.0&0.0\\\hline
BOIN-M&\underline{71.8}&23.2&4.7&0.2&0.0&0.0&\underline{21.8}&10.9&2.8&0.4&0.0&0.0\\\hline
Keyboard&\underline{75.4}&21.5&2.9&0.2&0.0&0.0&\underline{24.5}&9.2&2.0&0.3&0.0&0.0\\\hline
Keyboard-TS&\underline{73.2}&22.2&4.2&0.4&0.0&0.0&\underline{22.2}&10.2&3.0&0.6&0.1&0.0\\\hline
Keyboard-TS5\%&\underline{71.5}&24.5&3.8&0.2&0.0&0.0&\underline{21.8}&11.5&2.5&0.3&0.0&0.0\\\hline
Keyboard-G&\underline{74.8}&22.1&2.9&0.2&0.0&0.0&\underline{24.2}&9.9&1.8&0.2&0.0&0.0\\\hline
Keyboard-M&\underline{71.3}&24.2&4.4&0.1&0.0&0.0&\underline{21.8}&11.3&2.7&0.3&0.0&0.0\\\hline
\end{tabular}}
  \end{center}
\end{table}

\section{Discussion}\label{sec4}
We proposed multi-armed bandit BOIN and keyboard designs for dose-finding trials. Multi-armed bandits are very simple and powerful methods to determine actions to maximize a reward in a limited number of trials. Model-assisted designs have been actively discussed in recent years because of their simplicity and high performance. By applying the multi-armed bandits to model-assisted designs, we have derived simple and well-performing dose-finding designs. Among the multi-armed bandits, we deal with Thompson sampling which determines actions based on random samples from a posterior distribution. In addition, we applied regularized Thompson sampling-$\ep$ and greedy algorithm. The greedy algorithm determines a dose based on a posterior mean. In addition, we proposed a method to determine a dose based on a posterior median.

In the simulation studies, we confirmed that the PCMIs were better when there was a correct MTD above the middle doses. In actual dose-finding trials, low doses near the starting dose are a very safe dose (not assuming MTD) to set, and the MTD is assumed in medium doses and higher. Therefore, our proposed designs have superior PCMI in actual trials. We also confirmed that the number of patients treated MTD was higher than non multi-armed bandit designs. Hence, more data on the safety and efficacy on MTD can be collected.

Dose-finding trials for safety evaluation are conducted early phase in drug development and are very important trials to determine the maximum dose level for safety perspective. A failure in the identification of MTD will lead to failure in the late phase \cite{Conaway2019}, and the drug may not be able to be released to the public. We believe that our proposed designs is very useful.

\bigskip
\noindent

\bigskip
\noindent
{\bf Acknowledgments.}\ \ 
The author would like to thank Associate Professor Hisashi Noma for his encouragement and helpful suggestions.





\nocite{*}
\bibliography{manuscript}%

\appendix
\section{Supplemental Table}\label{App}
\begin{table}[h!]
  \begin{center}
\caption{Comparisons of different values of $\ep$ in Thompson sampling-$\ep$\label{STable1}}
{\tabcolsep=4.25pt
\begin{tabular}{|c|c|c|c|c|c|c|c|}\hline
Design &  & \multicolumn{6}{c|}{Scenarios}\\\hline
& & 1 & 2 & 3 & 4 & 5 & 6 \\\hline
BOIN-TS1\%&\%MTD&53.9&37.8&48.7&49.2&51.1&72.3\\\hline
BOIN-TS3\%&\%MTD&53.6&38.8&48.4&50.2&51.5&72.5\\\hline
BOIN-TS5\% &\%MTD&53.8&38.8&49.6&50.5&52.6&71.5\\\hline
BOIN-TS10\%&\%MTD&50.6&37.0&48.6&50.5&52.5&73.2\\\hline
Keyboard-TS1\%&\%MTD&33.7&30.9&45.4&48.6&51.9&71.4\\\hline
Keyboard-TS3\%&\%MTD&30.7&30.5&45.4&48.9&53.5&72.0\\\hline
Keyboard-TS5\% &\%MTD&31.4&30.0&45.8&51.2&54.3&71.5\\\hline
Keyboard-TS10\%&\%MTD&23.9&29.0&42.2&51.0&55.4&72.1\\\hline
\end{tabular}}
  \end{center}
\end{table}

\end{document}